\documentclass[twocolumn,printnumbers,amsmath,amssymb,showpacs]{revtex4}
\usepackage{graphicx}
\usepackage{color}

\begin{document}
\title{Equivalence of glass transition and colloidal glass transition in the hard-sphere limit}

\date{\today}

\author{Ning Xu$^{1}$}
\author{Thomas K. Haxton$^2$}
\author{Andrea J. Liu$^2$}
\author{Sidney R. Nagel$^3$}

\affiliation{$^1$Department of Physics, The Chinese University of Hong Kong, Hong Kong, People's Republic of China; $^2$Department of Physics and Astronomy, University of Pennsylvania, Philadelphia, PA 19104, USA; $^3$The James Franck Institute, The University of Chicago, Chicago IL, 60637, USA}

\begin{abstract}
We show that the slowing of the dynamics in simulations of several model glass-forming liquids is equivalent to the hard-sphere glass transition in the low-pressure limit.  In this limit, we find universal behavior of the relaxation time by collapsing molecular-dynamics data for all systems studied onto a single curve as a function of $T/p$, the ratio of the temperature to the pressure.   At higher pressures, there are deviations from this universal behavior that depend on the inter-particle potential, implying that additional physical processes must enter into the dynamics of glass-formation.
\end{abstract}


\maketitle
In many amorphous systems, particle dynamics can slow down so dramatically as a control parameter is varied that fluid behavior is suppressed and the system forms an apparently rigid solid.  One common example is a molecular liquid which, upon sufficient supercooling, becomes a glass with no ability to flow~\cite{glassreview}.  Here, the temperature $T$ governs, among other things, what energy barriers may be crossed as the system attempts to equilibrate~\cite{landscape}.  Another example is a colloidal fluid, in which micron-sized particles driven by Brownian motion form a colloidal glass when the density $\phi$, or pressure $p$, is raised~\cite{jammingbook,colloidreview}.  In this case, density controls the particle dynamics by governing the amount of free volume available for particles to rearrange~\cite{cohen}.  While the conceptual frameworks for understanding the slow dynamics in these two cases appear to focus on two very different physical effects, many aspects of these structural glass transitions appear to be similar~\cite{jammingbook}.   In neither case is it known whether an underlying phase transition accompanies glass formation; as the dynamics slow down, no obvious structural changes or unambiguous static order parameters emerge~\cite{glassreview,jammingbook}.  Without such a theoretical underpinning, it has been impossible to determine whether both phenomena are driven by the same physics or whether varying $T$ and $p$ slows the dynamics in intrinsically different ways.

By analyzing relaxation in several models that have frequently been used to capture different features of dynamical slowing, we are able to identify a universal aspect of glass formation that emerges at low pressures and temperatures.  In this limit, we can scale the relaxation times $\tau$ for all systems studied onto a single master plot as a function of the {\em single} variable $T/p$.  This has two important consequences. On the one hand, the scaling collapse allows one to see unambiguously that at sufficiently low pressures and temperatures, the colloidal glass transition and molecular glass transition are manifestations of the same phenomenon.  In this limit the temperature-driven molecular glass transition is {\em equivalent} to the pressure-driven hard-sphere colloidal glass transition; temperature and pressure are equally important for slowing relaxation processes.  On the other hand, deviations from this universal behavior away from this limit demonstrate that at least two distinct physical processes can enter to produce the slowdown of particle motion.

Our models consist of spheres with interparticle interactions $V(r_{ij})$, where $r_{ij}$ is the distance between particles $i$ and $j$ with diameters $\sigma_i$ and $\sigma_j$, respectively.  The particles do not interact at large separations: $V(r_{ij})=0$ for $r_{ij}\ge \sigma_{ij}$, where $\sigma_{ij}=(\sigma_i+\sigma_j)/2$.  If $r_{ij} < \sigma_{ij}$, the particles repel each other according to either the Weeks-Chandler-Andersen potential~\cite{wca}
\begin{equation}
V(r_{ij})=\frac{\epsilon}{72} \left[ \left(
\frac{\sigma_{ij}}{r_{ij}}\right)^{12} - 2 \left(
\frac{\sigma_{ij}}{r_{ij}}\right)^6 + 1\right],  \label{eq:WCA}
\end{equation}
or a simple power law
\begin{equation}
V(r_{ij})={\frac \epsilon \alpha} (1-r_{ij}/\sigma_{ij})^\alpha,  \label{eq:standardpot}
\end{equation}
where we use $\alpha=2$ (harmonic repulsion), $\alpha=5/2$ (Hertzian repulsion) or $\alpha=0$ (hard-sphere repulsion).
All of these systems have a jamming transition in the limit of zero temperature, $T \rightarrow 0$, and zero pressure, $p \rightarrow 0$~\cite{ohern03}.

We use molecular dynamics at fixed temperature and pressure to simulate three-dimensional systems composed of 50:50 mixtures of  $N=1000$ particles with diameters $\sigma$ and $\sigma_L=1.4 \sigma$ and the same mass, $m$. Temperature has units $\epsilon$, pressure has units $\epsilon/\sigma^3$, and time has units $\sqrt{m \sigma^2/\epsilon}$.  The Boltzmann constant, $k_B$, is set to unity.  We use an event-driven code for the hard-sphere simulations.  Periodic boundary conditions are applied in all directions.  We calculate the self-part of the intermediate structure factor: $S(\vec{k},t)=\frac{2}{N}\sum_i {\rm exp}(i\vec{k}\cdot [\vec{r}_i(t)-\vec{r}_i(0)])$~\cite{kobandersen}, where the sum is over all large particles, $\vec{r}_i(t)$ is the location of particle $i$ at time $t$, and $\vec{k}$ is chosen in the $x-$direction.  The amplitude of $\vec{k}$ satisfies the periodic boundary conditions and is approximately the value at the first peak of the static structure factor.  We define the relaxation time $\tau$ to be the time at which $S(\vec{k},\tau)=e^{-1}S(\vec{k},0)$.  We take data after a system has been equilibrated for several $\tau$.

\begin{figure}
\includegraphics[width=0.35\textwidth]{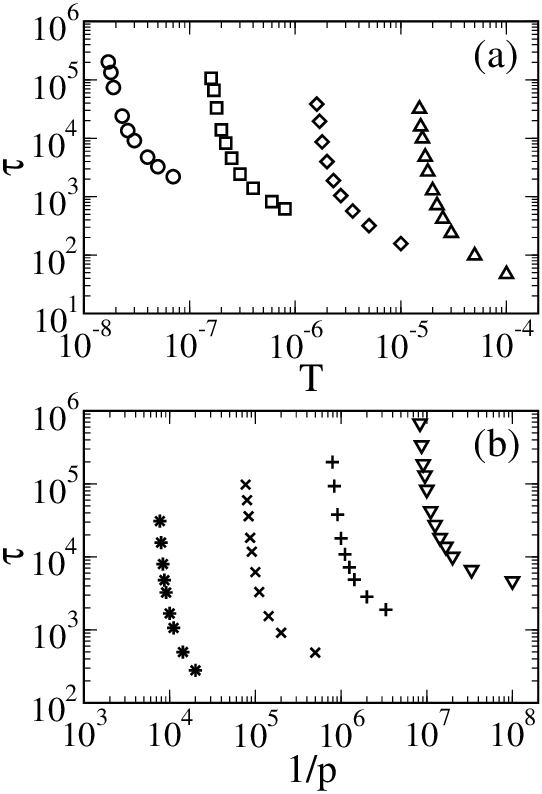}
\caption{\label{fig:unscaled}
 Relaxation time versus different control parameters for a system of 1000 particles with harmonic repulsions.  (a)  Relaxation time $\tau$ versus temperature $T$ at different fixed pressures, $p$ : $p=2 \times 10^{-7}$ (circles), $p=2 \times 10^{-6}$ (squares), $p=2 \times 10^{-5}$ (diamonds), and $p=2 \times 10^{-4}$ (upward triangles). (b)  Relaxation time $\tau$ versus inverse pressure $1/p$ at different fixed temperatures: $T=10^{-8}$ (downward triangles), $T=10^{-7}$ (pluses), $T=10^{-6}$ (crosses), and $T=10^{-5}$ (stars).
}
\end{figure}

Fig.~\ref{fig:unscaled}(a) shows the relaxation time $\tau$ versus temperature $T$ for a system with harmonic repulsions ($\alpha=2$ in Eq.~\ref{eq:standardpot}) that is cooled at different fixed low pressures, $p$.  This corresponds to the standard trajectory for experiments on supercooled liquids undergoing the glass transition, a trajectory in which $p$ is typically fixed at atmospheric pressure.  Fig.~\ref{fig:unscaled}(b) shows $\tau$ versus $1/p$ at different fixed $T$ for the same system; in these trajectories, we raise $p$ at fixed $T$ as is typically done in experiments on colloidal systems.  As expected, $\tau$ increases with decreasing $T$ and increasing $p$.

\begin{figure}
\includegraphics[width=0.38\textwidth]{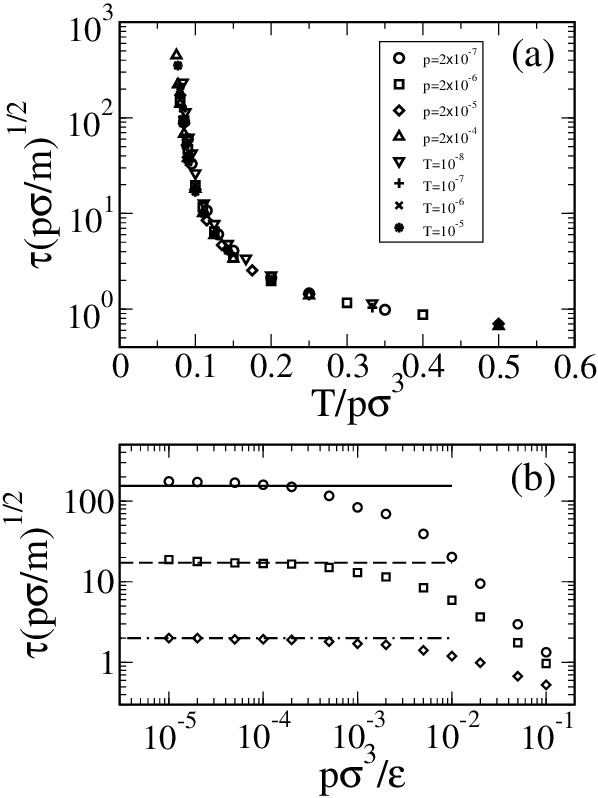}
\caption{\label{fig:scaled}
  Collapse of all the relaxation time data shown in Fig.~\ref{fig:unscaled}.  (a)  Scaled relaxation time, $\tau/\sqrt{m/p\sigma}$, versus scaled temperature $T/p \sigma^3$.    (b) Scaled relaxation time $\tau/\sqrt{m/p\sigma}$ versus scaled pressure $p \sigma^3/\epsilon$, for  $T/p \sigma^3= 0.08$ (circles),  $T/p\sigma^3=0.1$ (squares) and $T/p\sigma^3=0.2$ (diamonds).  Horizontal lines show the limiting values of $\tau \sqrt{p \sigma/m}$ as $p \sigma^3/\epsilon \rightarrow 0$.
}
\end{figure}

Fig.~\ref{fig:scaled}(a) shows that we can collapse the data in Fig.~\ref{fig:unscaled} for {\em all} the trajectories, both at fixed pressure and fixed temperature, onto a single curve by scaling the relaxation time by $\sqrt{m/p \sigma}$ and temperature by $p\sigma^3$, so that
\begin{equation}
\tau/\sqrt{m/p \sigma} = F \bigl (T/p \sigma^3 \bigr ), \label{eq:dimb}
\end{equation}
This collapses data ranging over 4 decades of temperature and pressure.  This remarkable collapse is achieved by using the time scale, $\sqrt{m/p \sigma}$, and energy scale, $p \sigma^3$,  to make relaxation time and temperature dimensionless~\cite{dacruz}. The characteristic time $\sqrt{m/p \sigma}$ is proportional to the time for a particle starting at rest to move its diameter $\sigma$ due to a pressure $p$ and is the duration of a pressure-driven particle rearrangement.  The dependence on $T/p\sigma^3$ implies that both $T$ and $p$ are equally important in controlling the dimensionless relaxation time.

Dimensional analysis provides a starting point for understanding the implications of the data collapse of Eq.~\ref{eq:dimb}.  The dimensionless time $\tau/\sqrt{m/p \sigma}$ can be written as a function of the dimensionless variables of the system.  In addition to $T/p\sigma^3$, another dimensionless ratio is $p \sigma^3/\epsilon$, where $\epsilon$ sets the scale of the interaction energy in Eq.~\ref{eq:standardpot}.  There are no other independent dimensionless variables for the system with harmonic repulsions.  The dimensionless relaxation time must therefore satisfy
\begin{equation}
\tau \sqrt{p \sigma/m} = f \bigl (T/p \sigma^3, p \sigma^3/\epsilon \bigr ). \label{eq:dima}
\end{equation}

The data shown in Fig.~\ref{fig:scaled}(a) all lie at low pressures, where the second argument in Eq.~\ref{eq:dima}, $p\sigma^3/\epsilon$, is small.  In Fig.~\ref{fig:scaled}(b), we show the scaled relaxation time versus $p \sigma^3/\epsilon$, at three values of $T/p \sigma^3$.  In all cases, the data approach an asymptotic value at low $p \sigma^3/\epsilon$.  Thus, in the low-$p \sigma^3/\epsilon$ limit, $\tau \sqrt{p\sigma/m}$ is a function of $T/p\sigma^3$ only, consistent with the collapse of Fig.~\ref{fig:scaled}(a).

The limit, $p \sigma^3/\epsilon \rightarrow 0$, is always satisfied in the hard-sphere limit, $\epsilon \rightarrow \infty$.    Thus, the relaxation time for hard spheres should collapse onto the same scaling form as in Fig.~\ref{fig:scaled}.  Moreover, all potentials that behave as hard spheres in the low-pressure limit by preventing overlap between particles---namely, all potentials with finite-ranged repulsions--should also collapse onto the same form at low $p \sigma^3/\epsilon$.  This is not what we would expect from dimensional analysis alone, since different potentials can contain additional dimensionless parameters, such as the exponent $\alpha$ in Eq.~\ref{eq:standardpot}.  However, the physics of the $p\sigma^3/\epsilon \rightarrow 0$, hard-sphere limit suggests that collapse should occur, irrespective of the value of $\alpha$, as long as $\alpha \ge 0$ so that the potential is repulsive.

\begin{figure}
\includegraphics[width=0.38\textwidth]{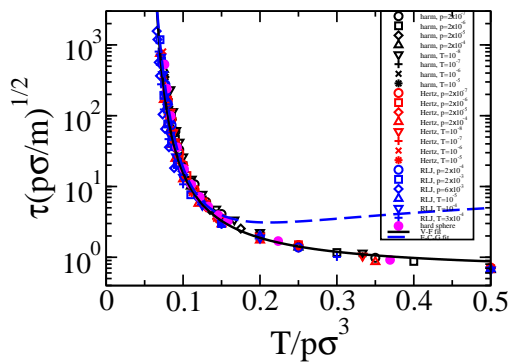}
\caption{\label{fig:allscaled}
(Color online) Scaled relaxation time, $\tau/\sqrt{m/p\sigma}$, versus scaled temperature $T/p \sigma^3$ for all the data for the harmonic potential (black) in Fig.~\ref{fig:unscaled} as well as for the Hertzian potential, $\alpha=5/2$ (red), the hard-sphere potential (magenta), and the Weeks-Chandler-Andersen potential (blue).  Black solid curve is the Vogel-Fulcher fit: $y=0.59 \exp(0.18/(x-0.045))$, where $y=\tau/\sqrt{m/p\sigma}$ and $x=T/p\sigma^3$.  Blue-dashed curve is a fit to the Elmatad-Chandler-Garrahan form: $y=3.1 \exp[0.064(x^{-1}-4.72)^2]$.
}
\end{figure}

This is corroborated in Fig.~\ref{fig:allscaled}, where we show the data of Figs.~\ref{fig:unscaled} and \ref{fig:scaled} for harmonic repulsions ($\alpha=2$ in Eq.~\ref{eq:standardpot}) together with data for three other potentials:  Hertzian repulsions ($\alpha=5/2$ in Eq.~\ref{eq:standardpot}), the hard-sphere potential ($\alpha=0$ in Eq.~\ref{eq:standardpot}), and the Weeks-Chandler-Andersen potential (Eq.~\ref{eq:WCA}).   Indeed, all of these systems collapse onto the {\it same scaling form}.  Thus, for the low-pressure (low $p \sigma^3/\epsilon$) data considered here, $\tau \sqrt{p\sigma/m}$ is a function of $T/p\sigma^3$ only and does not depend separately on the interaction potential - either on its form ({\em i.e.}, whether it is given by Eq.~\ref{eq:WCA} or ~\ref{eq:standardpot} with different exponents $\alpha$) or on its overall magnitude, $\epsilon$.  This represents a major simplification of the relaxation-time data.

While the scaled dynamics are independent of dimensionless parameters characterizing the inter-particle potential, they may be affected by other dimensionless numbers characterizing the system.  For example, relaxation time data for hard-sphere systems with different polydispersities or diameter ratios will not necessarily collapse.

The data collapse suggests a way of looking at the glass transition in this limit where the relaxation time is only a function of $T/p \sigma^3$.  The ratio of $T/p$ corresponds to an effective volume created by using thermal energy to do work against the pressure: $p \Delta V \sim T$.  In the hard-sphere glass transition, the system relaxes via free volume.  Likewise, soft-sphere liquids can create free volume by using thermal energy.  At high temperatures, where there is plenty of this ``thermal" free volume, the system relaxes rapidly; at low temperatures, where there is less thermal free volume, the system relaxes more slowly.

The form of the scaling function, $F(x)$, shown in Fig.~\ref{fig:allscaled}, should tell us whether the system has a thermodynamic glass transition.  If such a transition exists, as has been suggested theoretically for hard spheres~\cite{zamponi}, then $F(x)$ should diverge at a nonzero value of $T/p \sigma^3$, {\em i.e.}, at nonzero $T$ for soft-sphere liquids or at finite $p$ for hard spheres.  Fig.~\ref{fig:allscaled} shows that over the dynamic range of our simulations, the Vogel-Fulcher form,
\begin{equation}
\tau/\sqrt{m/p \sigma}=C \exp \bigl (A/(x-x_0) \bigr ) \label{eq:VF}
\end{equation}
with $x=T/p\sigma^3$, provides a reasonable fit (solid line) with $x_0=0.045$, $A=0.18$ and $C=0.59$.  The scaling collapse and Eq.~\ref{eq:VF} imply that $T_0/p\sigma^3=x_0$ so that $T_0=p \sigma^3 x_0$.  In other words, $T_0$ increases with pressure.  These results are consistent with recent numerical studies on spheres with harmonic repulsions~\cite{berthier}, which show that $T_0$ increases with packing fraction above some critical value.  The parameter $T_0$ is often used to parameterize the fragility of the system~\cite{angell,angellrev}; thus, the fit to Eq.~\ref{eq:VF} implies that fragility increases with pressure.

\begin{figure}
\includegraphics[width=0.38\textwidth]{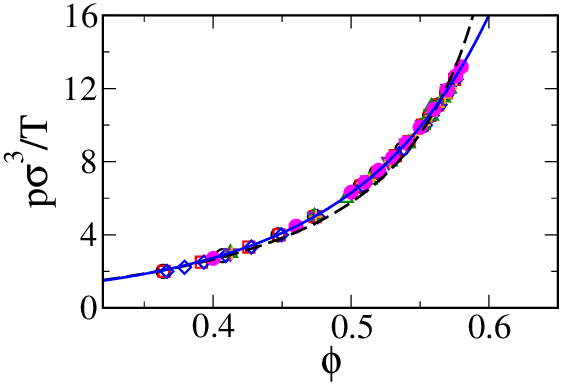}
\caption{\label{fig:eos} (Color Online)  Equation of state for three-dimensional systems at low $p \sigma^3/\epsilon$ for hard sphere systems (solid magenta circles) and systems with harmonic repulsions measured along both constant temperature and constant pressure trajectories.  The dashed line is a fit to free-volume theory, $p \sigma^3/T=0.98\phi/(1-(\phi/\phi_c)^{1/3})$, with $\phi_c=0.66$.  The solid line is an empirical fit given by $p \sigma^3/T=\phi/(1-(\phi/1.13))^{4.35}$.}
\end{figure}

We also find that a fitting form proposed by Elmatad, et al.~\cite{elmatad}, for which $\tau/\sqrt{m/p \sigma}$ diverges only at $T/p \sigma^3=0$, provides an equally good fit in the regime of interest at small $T/p \sigma^3$.  For this fitting form,
\begin{equation}
\tau/\sqrt{m/p \sigma}=C_1 \exp \biggl [ A_1^2 \bigl (\frac{1}{x}-\frac{1}{x_1} \bigr ) ^2 \biggr ], \label{eq:ECG}
\end{equation}
where $x=T/p\sigma^3$.  We find $x_1=0.21$, $A_1=0.25$ and $C_1=3.1$.  Thus, we cannot distinguish whether the scaling function diverges at nonzero or zero $T/p\sigma^3$.  This is not surprising, since even experiments with 17 decades of dynamic range cannot tell whether there is a thermodynamic glass transition.  However, we note that if there is no thermodynamic glass transition, the scaling function diverges at $T/p\sigma^3=0$, so that the relaxation time diverges in the double limit $T/p\sigma^3  \rightarrow 0$, $p\sigma^3/\epsilon \rightarrow 0$.  This double limit corresponds to the zero-temperature jamming transition of frictionless spheres with finite-ranged repulsions, also known as Point J~\cite{ohern03}.  This implies that the glass transition is controlled by Point J if there is no intervening thermodynamic glass transition.

The relaxation time is not the only quantity to exhibit data collapse in the low $p\sigma^3/\epsilon$ limit.  Other unrelated quantities, including those that are independent of the dynamics, such as the packing fraction, $\phi$, should also exhibit collapse in that limit.  The packing fraction is the number density made dimensionless by the average particle volume.  For a fixed potential, $\phi$ must be
expressible as a function of $T/p \sigma^3$ and $p \sigma^3/\epsilon$.  In the low-pressure, hard-sphere limit where $p \sigma^3/\epsilon \ll 1$, it should satisfy $\phi=\tilde H \bigl (T/p \sigma^3\bigr )$ with the same scaling function $\tilde H(x)$ for all finite-ranged repulsive potentials.  This function can be inverted to yield the equation of state
\begin{equation}
p \sigma^3/T= H \bigl (\phi). \label{eq:eosb}
\end{equation}
Fig.~\ref{fig:eos} shows the data collapse for the equation of state along different trajectories and for different potentials for our bi-disperse systems in three dimensions. 
We have shown two fits to the data.  The dashed line is a fit to free volume theory for $\phi$ near the fitting parameter $\phi_c=0.66$.  This form fits experimental data for colloidal hard spheres and numerical data for hard spheres reasonably well~\cite{randy}, but is clearly unsatisfactory here.   The solid line is an empirical fit to the data  (see caption).

Our results show that the data for the relaxation time in many different systems collapse onto a single curve of $\tau \sqrt{p\sigma/m}$ versus $T/p\sigma^3$. This collapse, however, is confined to $p \ll \epsilon/\sigma^3$.  We find that upon increasing the pressure, there are deviations from the scaling collapse (see Fig.~\ref{fig:scaled}(b)).  This implies that there are additional contributions to relaxation near the glass transition, beyond the particle rearrangements facilitated by the thermal free volume.  At large $p\sigma^3/\epsilon$, relaxation also occurs because thermal fluctuations drive the system across energy barriers.  This contribution could perhaps be taken into account by introducing an effective pressure-dependent radius for the particles (such as the Barker-Henderson radius~\cite{barker}).
As $p\sigma^3/\epsilon$ increases at fixed $T/p\sigma^3$ and the density of the system increases above the jamming point, the effective radius shrinks, increasing the effective free volume beyond the thermal free volume created by the thermal energy working against the pressure.

The relative effects of $T$ and $p$ on $\tau$ in glass-forming liquids have been studied experimentally, with differing conclusions~\cite{tarjuskivelson,winmenon}.  In order to understand the consequences for molecular liquids, we must first understand the corrections to the leading hard-sphere behavior when $p \sigma^3/\epsilon$ is no longer small.  Molecular liquids typically have densities high compared to that of a system near the jamming threshold.  They also have long-ranged attractions as well as the short-ranged repulsions considered here.  For these glass-forming liquids, hard spheres may still be a useful starting point, but at least one other distinct contribution to the relaxation must also be considered.

We thank N. Menon for instructive discussions.  This work was supported by DE-FG02-05ER46199 (AL, NX, and TH), DE-FG02-03ER46088 (SN and NX), and NSF-MRSEC DMR-0820054 (SN).

\end{document}